\documentclass[a4paper,12pt]{article}
\usepackage[pctex32]{graphics}

\begin{document}
\topmargin 0pt \oddsidemargin 0mm

\renewcommand{\thefootnote}{\fnsymbol{footnote}}
\begin{titlepage}

\vspace{5mm}
\begin{center}
{\Large \bf Origin of holographic dark energy models}
\vspace{12mm}

{\large    Yun Soo Myung\footnote{e-mail
 address: ysmyung@inje.ac.kr} and Min-Gyun Seo}
 \\
\vspace{10mm} {\em  Institute of Mathematical Science and School
of Computer Aided Science \\ Inje University, Gimhae 621-749,
Korea }

\end{center}

\vspace{5mm} \centerline{{\bf{Abstract}}}
 \vspace{5mm}

We investigate the origin of holographic  dark energy models which
were recently proposed to explain the dark energy-dominated
universe. For this purpose, we introduce the spacetime foam
uncertainty of $\delta l \ge l_{\rm p}^{\alpha}l^{\alpha-1}$. It
was argued that the case of $\alpha=2/3$ could describe the dark
energy with infinite statistics, while the case of $\alpha=1/2$
can describe the ordinary matter with Bose-Fermi statistics.
However, two cases may lead to the holographic energy density if
the latter recovers from the geometric mean of UV and IR scales.
Hence the dark energy with infinite statistics based on the
entropy bound is not an ingredient for deriving  the holographic
dark energy model. Furthermore, it is shown that the agegraphic
dark energy models are the holographic dark energy model with
different IR length scales.

\end{titlepage}
\newpage
\renewcommand{\thefootnote}{\arabic{footnote}}
\setcounter{footnote}{0} \setcounter{page}{2}

\section{Introduction}
Observations of supernova type Ia suggest that our universe is
accelerating~\cite{SN}. Considering the ${\rm \Lambda}$CDM
model~\cite{SDSS,Wmap1}, the dark energy and cold dark matter
contribute $\Omega^{\rm ob}_{\rm \Lambda}\simeq 0.74$ and
$\Omega^{\rm ob}_{\rm CDM}\simeq 0.22$ to the critical density of
the present universe. Recently, the combination of WMAP3 and
Supernova Legacy Survey data shows a significant constraint on the
equation of state (EOS) for the dark energy, $w_{\rm
ob}=-0.97^{+0.07}_{-0.09}$ in a flat universe~\cite{WMAP3,SSM}.

Although there exist a number of dark energy models~\cite{CST},
the two promising candidates are the cosmological constant  and
the quintessence scenario~\cite{UIS}. The EOS for the latter is
determined dynamically by the scalar or tachyon.

On the other hand, there exist interesting models of the dynamical
dark energy which satisfy  the holographic principle but have
different origins. One is the holographic dark energy
model~\cite{LI,HM} and the other is the agegraphic dark energy
model~\cite{CAI}. The first is derived from the energy
bound~\cite{CKN,myung}, while the latter is based on the
K\'{a}rolyh\'{a}zy relation of quantum fluctuations of
time~\cite{Karo,ND,Sas} and the time-energy
uncertainty~\cite{Maz}. It seems that the agegraphic dark energy
density is clearly understood  because its energy is just  the
minimum energy of
 spacetime fluctuations derived from the time-energy uncertainty.
 However, the origin of  holographic energy density
 remains unclear and obscure because it was obtained from the energy bound using the black hole.

Recently, Ng\cite{Ng1} has proposed that the entropy bound is
designed for deriving the holographic dark energy. On the other
hand,  the energy bound was  used for describing the holographic
dark energy~\cite{CKN,LI}.  Hence, it is necessary to reexamine
the holographic dark energy model based on the energy bound.

In this Letter, we address this issue and  explore the connection
between holographic and agegraphic dark energy models.

\section{Spacetime foam uncertainty}

We start with reviewing holographic dark energy model.
 This model  comes from the energy bound~\cite{CKN,myung}
\begin{equation}  \label{EB}
 E_{\rm \Lambda} \le E_{BH} \to
l^3 \rho_{\rm \Lambda}\le m_{\rm p}^2l,
\end{equation}
 where the
vacuum energy density is given by $\rho_{\rm \Lambda}=\Lambda^4$
with the UV cutoff $\Lambda$ and $l$  is the length scale (IR
cutoff) of the system. Choosing the saturation of this bound leads
to holographic  energy density
\begin{equation} \label{hed}
\rho_{\rm \Lambda} \sim \frac{m_{\rm p}^2}{l^2}\sim
\frac{1}{(l_{\rm p}l)^2}
\end{equation}
We note that the energy bound Eq.(\ref{EB}) implies another
entropy bound,
\begin{equation}  \label{ENBa}
S_{\rm \Lambda}\le \Big(m_{\rm p}^2A\Big)^{3/4}
\end{equation}
with $A=4\pi l^2$ is the area of system. This is not a covariant
entropy  bound.

 On the other hand,  the agegraphic dark energy model
is based on the K\'{a}rolyh\'{a}zy relation of quantum
fluctuations of time~\cite{Karo,ND,Sas}
\begin{equation}
 \delta t=\lambda t^{2/3}_{\rm p}t^{1/3}
\end{equation}
 and
the time-energy uncertainty
\begin{equation} \label{TEU} \Delta E \sim t^{-1}
\end{equation} in the
Minkowiski spacetime. This  gives us the agegraphic  energy
density~\cite{Maz}
\begin{equation}  \label{aed}
 \rho_{\rm T}\sim
\frac{\Delta E}{(\delta t)^3}\sim \frac{m_{\rm p}^2}{t^2}.
\end{equation}

Furthermore, Ng has proposed the spacetime foam model where the
covariant entropy bound,
\begin{equation}  \label{ENB}
S_{\rm \Lambda}=\Lambda^3l^3 \le S_{BH}= m_{\rm p}^2l^2 \sim A
\end{equation} plays a crucial role
for conjecturing the presence of  the holographic  energy density
in Eq.(\ref{hed}).  A basic feature of this model comes from the
spacetime foam uncertainty~\cite{Ng1,Ng2,AKN}
\begin{equation}
\delta l \ge l_{\rm p}^{\alpha}l^{\alpha-1}. \end{equation}
 Explicitly, the $\alpha=2/3$ case of
 holographic uncertainty   could describe the dark
energy with infinite statistics, while the $\alpha=1/2$ case of
random-walk uncertainty  can describe the ordinary matter with
Bose-Fermi statistics. The important properties are summarized in
Table I.
\begin{table}
\caption{Summary of spacetime foam (STF) model~\cite{Ng1}. Here HM
(RWM) denote holographic (random-walk) models, and B/F represent
Bose-Einstein/Fermi-Dirac statistics. $A$ is the area of system. }
\begin{tabular}{|c|c|c|c|c|}
  \hline
  STF model   & distance fluctuations & entropy bound  &energy/matter& statistics \\
  \hline
  HM   & $\delta l \ge (l_{\rm p}^2l)^{1/3}$ & $A$ & dark energy & infinite\\
 RWM  & $\delta l \ge (l_{\rm p}l)^{1/2}$& $A^{3/4}$& ordinary matter& B/F \\
  \hline
\end{tabular}
\end{table}

We are in a position to point out the connection between
holographic dark energy  and spacetime foam models. Assuming the
relation between the UV cutoff and distance uncertainty
\begin{equation}
\Lambda \sim \frac{1}{\delta l},
\end{equation}
we derive the holographic uncertainty from the entropy bound
Eq.(\ref{ENB})~\cite{Ng3} as
\begin{equation} \label{EHM}
S_{\Lambda} \le S_{BH} \to \delta l\ge (l_{\rm p}^2l)^{1/3}
\end{equation}
and  the random-walk uncertainty from the energy bound
Eq.(\ref{EB}) as
\begin{equation} \label{ERM}
E_{\Lambda} \le E_{BH} \to \delta l\ge (l_{\rm p}l)^{1/2}.
\end{equation}
The above is  reasonable because the UV cutoff usually determines
the minimal detectable length~\cite{CX}. Hence it seems that the
entropy bound (energy bound) are closely related to the HM (RWM),
respectively.

Now we wish to obtain the holographic energy density $\rho_{\rm
\Lambda}$ from the entropy bound of Eq.(\ref{EHM}) and the energy
uncertainty. For this purpose, we introduce  the delocalized
states which  have typical Heisenberg energy \footnote{Actually,
there are two approaches: bulk holography and holographic
screens~\cite{Thom}. Two are closely related to each other as the
UV-IR connection. In the bulk holographic approach, it is natural
to postulate that uniformly distributed bulk holographic degrees
of freedom are delocalized on the size $l$ of the system. Then,
the Heisenberg quantum energy of each delocalized  holographic
degrees of freedom is $ E^{\rm B}_{\rm del} \sim 1/l$ with
$\hbar=1$.  In this case, the quantum contribution to the global
vacuum energy density is given by $\delta \Lambda_4 \sim
 E^{\rm B}_{\rm del}\frac{ N_{\rm sur}}{l^3}$. The total number of degrees of
freedom $ N_{\rm sur}=\frac{l^2}{l^2_{\rm p}}$ is determined by
the gravitational holography. Here we observe an important
relation between bulk holographic and spacetime foam approaches:
$\frac{ N_{\rm sur}}{l^3} =\frac{1}{l^2_{\rm p}l}=\frac{1}{(\delta
l)^3 }$. Consequently, one finds  $\delta
\Lambda_4=\rho_{\Lambda}$. }

\begin{equation}
  E^{\rm B}_{\rm del}
\sim \frac{1}{l} \end{equation}
 in the bulk~\cite{Thom}. In this case,   the Bekenstein-Hawking entropy $S_{BH}=N_{\rm sur}$  takes
into account the gravitational holography properly.
 Then we obtain the relation
\begin{equation} \label{rhohm}
\rho_{\rm HM}=\frac{ E^{\rm B}_{\rm del}}{(\delta
l)^3}=\frac{m_{\rm p}^2}{l^2} \sim \rho_{\rm \Lambda}
\end{equation}
which shows that  the holographic energy density could be derived
from the covariant entropy bound and  the spacetime fluctuations.
Here we mention that Eq.(\ref{rhohm}) is consistent with the
holographic model proposed in Ref.\cite{Ng2}.

We  check  that $ E^{\rm B}_{\rm del}$  may fit into a UV cell of
size $l_{\rm p}$ in the holographic screen  ($ E^{\rm S}_{\rm del}
\sim 1/l_{\rm p})$ as well.  For this purpose,  we consider the
system which is composed of $N$ UV cells~\cite{Pad}. Each cell has
a Poissonian fluctuation in energy of amount $E_{\rm p}\sim
1/l_{\rm p}$. Then the root-mean-square fluctuation of energy will
be
\begin{equation}
\Delta E_{\rm Po}=\sqrt{<(\Delta E_{\rm Po})^2>}=
\frac{\sqrt{N}}{l_{\rm p}}
\end{equation}
which fits into the UV cell but it is proportional to  the factor
$\sqrt{N}$.
 This  provides the energy density
\begin{equation} \label{rhohmc1}
\rho_{\rm Po}=\frac{\Delta E_{\rm Po}}{l^3}=\frac{\sqrt{N}}{l_{\rm
p}l^3}.
\end{equation}
Choosing $N=N_{\rm sur}$, we arrive at  the holographic energy
density
\begin{equation} \label{rhohmc2}
\rho_{\rm Po}=\frac{\sqrt{N_{\rm sur}}}{l_{\rm
p}l^3}=\frac{1}{l_{\rm p}^2l^2}.
\end{equation}
Hence, we observe that $\sqrt{N_{\rm sur}}E^{\rm B}_{\rm del}$
fits into a UV cell on the holographic screen as
\begin{equation} \label{rhohmc3}
\sqrt{N_{\rm sur}}E^{\rm B}_{\rm del} \simeq \frac{1}{l_{\rm p}}
\sim E^{\rm S}_{\rm del}.
\end{equation}
This indicates how  IR fluctuations in the bulk can fit into
 UV cells on the screen. Also the conversion factor $\sqrt{N_{\rm sur}}$ could be easily explained by
introducing  a screen-bulk redshift factor of $1/\sqrt{g_{00}}$.
The apparent horizon is a surface of infinite redshift, so a
regulated screen must be employed. The Planck length and energy
may be taken as UV cutoffs of local screen degrees of freedom. In
this case, the inverse of screen-bulk  redshift factor,
\begin{equation}\sqrt{g_{00}} \sim \frac{l_{\rm p}}{l} \sim
\frac{1}{\sqrt{N_{\rm sur}}}\end{equation} gives a bulk quantum
energy $E^{\rm B}_{\rm del}$. The definite connection is given by
UV-IR connection as~\cite{GPP,myungent}
\begin{equation}
E^{\rm B}_{\rm del}=\sqrt{g_{00}} E^{\rm S}_{\rm del}
\end{equation}
which confirm Eq.(\ref{rhohmc3}) clearly.

 On the other hand, from the energy bound of Eq.(\ref{ERM}),
it seems  difficult to derive the holographic energy density
because we do not know a form of $\Delta E_{\rm RWM}$ as
\begin{equation}
\rho_{\rm RWM}=\frac{\Delta E_{\rm RWM}}{(\delta
l)^3}=\frac{\Delta E_{\rm RWM}}{(l_{\rm p}l)^{3/2}}.
\end{equation}
Assuming that  $\Delta E_{\rm RWM}\sim 1/\sqrt{l_{\rm
p}l}$~\footnote{According to Ref.\cite{Ng1}, $\rho_{\rm RWM}$ is
bounded between $(ll_{\rm p})^{-2}$ and $l^{-5/2}l_{\rm
p}^{-3/2}$. Hence, this assumption is likely to be accepted.}, one
finds that $\rho_{\rm RWM} \sim \rho_{\rm \Lambda}$. However, it
is  unclear why the energy  of spacetime fluctuations is inversely
proportional to the geometric mean of $\sqrt{l_{\rm p}l}$ of
distance $l$ and Planck length $l_{\rm p}$ when the energy bound
is working for ordinary matter. In order to explain this, we
introduce  two length scales $l_{\rm UV}=l_{\rm p}$ and $l_{\rm
IR}=l$ by assuming  that there is no connection between them. This
means that we will not introduce any bound. Two energy densities
of UV and IR scales are given by~\cite{Pad}
\begin{equation}
\rho_{\rm UV}=\frac{1}{l^4_{\rm p}}~~{\rm and}~~\rho_{\rm
IR}=\frac{1}{l^4}.
\end{equation}
Here $\rho_{\rm UV}$ determines the highest possible energy
density in the universe, while $\rho_{\rm IR}$ determines the
lowest possible energy density.
 Then the geometric
mean (GM) of two energy densities takes the form
\begin{equation}
\rho_{\rm GM}=\sqrt{\rho_{\rm UV}\rho_{\rm IR}}=\frac{1}{l^2_{\rm
p}l^2}
\end{equation}
which is just a form of holographic energy density $\rho_{\rm
\Lambda}$.  Importantly, the geometric mean\footnote{For
comparison, we introduce two others: average (mean)=$\frac{l_{\rm
p}+l}{2} \sim l$ and harmonic mean=$\frac{l_{\rm p} l}{2(l_{\rm p}
+l)}\sim l_{\rm p}$ for $l_{\rm p} \ll l$. Hence, for $l_{\rm p}
\ll l$, the relevant scale is the geometric mean. } of two length
scales leads to the minimum length of the RWM: $l_{\rm
GM}=\sqrt{l_{\rm UV}l_{\rm IR}}=\sqrt{l_{\rm p}l} \to \delta l$.
Consequently, the geometric mean of two energies leads to
\begin{equation}
E_{\rm GM}=\sqrt{(l_{\rm UV}^3\rho_{\rm UV})( l^3_{\rm
IR}\rho_{\rm IR})}=\frac{1}{\sqrt{l_{\rm p}l}}.
\end{equation}
This is what we expect to obtain for the energy for the RWM. That
is, if $\Delta E_{\rm RWM}=E_{ \rm GM}$, one could obtain the
holographic energy density from the RWM which is known to describe
the ordinary matter. At this time,  we do not prove that the
presumed proposition of $\Delta E_{\rm RWM}=E_{ \rm GM}$ is
correct. Here we  could support this by the dimensional argument.

We mention  cumulative effects of spacetime
fluctuations~\cite{Ng1}. If successive fluctuations are completely
anti-correlated (negative correlation: NC), the fluctuation
distance $\delta l$ is given by $l_{\rm UV}= l_{\rm p}$, being
independent of the size of distance $l$. If successive
fluctuations are completely correlated (positive correlation: PC),
the fluctuation distance $\delta l$ is given by $l_{\rm IR}= l$,
the size of distance $l$. The zero correlation (ZC) corresponds to
the RWM of  $\delta l \sim \sqrt{l_{\rm p} l}$, while the order of
correlation ($\delta l \sim (l_{\rm p}^2 l)^{1/3}$) for the HM is
between NC and ZC. This implies that the effects of quantum
gravity is strongest for UV (NC), while the effects of quantum
gravity is zero for the RWM (ZC). We remind the reader that the
quantum-gravitational effects of HM is between the strongest one
and zero.
\begin{center}
$\underline{l_{\rm p}--(l_{\rm p}^2l)^{1/3}---\sqrt{l_{\rm p}
l}---------- l}$

 ${\rm UV(NC)}--{\rm HM}--{\rm
RWM(ZC,GM)}------{\rm IR(PC)}$
\end{center}
The holographic uncertainty for  the entropy bound leads to the
holographic energy density. If this is unique, the entropy bound
should be used for describing the system including
self-gravitating effects only. Along this direction, we note that
for the HM, the individual fluctuations cannot be completely
random, as opposed to the no correlation of RWM. Hence, successive
fluctuations appeared to be entangled and somewhat anti-correlated
as a result of effects of quantum gravity. On the other hand, the
random-walk uncertainty for  the energy bound  could provide the
holographic energy density by choosing $\Delta E_{\rm RWM}=E_{\rm
GM}$. However, we do not know a close connection between RWM and
GM

In addition, different sources may lead to  the holographic energy
density. These are vacuum fluctuation energy~\cite{Pad},
entanglement entropy (energy), and Casimir
energy~\cite{myungent,LKL}.  Until now, there is no unique way to
give the holographic energy density.

\section{Holographic and agegraphic dark energy models}
Even though we got the holographic energy density, it is not
guaranteed that the holographic energy density could describe the
present accelerating universe. Here we choose
\begin{equation}
\rho_{\rm \Lambda}=\frac{3c^2 m^2_{\rm p}}{L^2}
\end{equation}
with a parameter $c$.
 In order for the holographic energy
density to describe the accelerating universe, we have to choose
an appropriate IR cutoff $L$. For this purpose, we may introduce
three length scales of the universe: the apparent horizon=Hubble
horizon for flat universe, particle horizon, and future event
horizon. The equation of state is defined by
\begin{equation} \label{EOS}
w_{\rm i}=-1-\frac{a}{3 \rho_{\rm i}}\frac{d\rho_{\rm i}}{da}
\end{equation}
 with the scale factor $a$. For the presence of interaction between two matters,
 one may introduce either the native EOS~\cite{WGA} or the effective EOS~\cite{KLM}. The density parameter is defined
by \begin{equation} \Omega_{\rm i}=\frac{\rho_{\rm i}}{3m_{\rm
p}^2 H^2}=\Big(\frac{c}{HL_{\rm i}}\Big)^2.\end{equation} Its
evolution is determined by
\begin{equation} \label{omee}
\frac{d\Omega_{\rm i}}{dx}=-3w_{\rm i}\Omega_{\rm i}(1-\Omega_{\rm
i})\end{equation} for the presence of $\rho_{\rm i}$ and the cold
dark matter (CDM) $\rho_{\rm m}$ with $x=\ln a$.

 When the CDM is present, the Hubble
horizon $L_{\rm HH}=1/H$ does not describe the accelerating
universe because its equation of state $w_{\rm HH}=0$ is the same
as the CDM dose. Using the first Friedmann equation with
$\rho_{\rm HH}=3c^2m^2_{\rm p}H^2$ leads to $(1-c^2)H^2=\rho_{\rm
m}/3m^2_{\rm p}$ with $\rho_{\rm m}=\rho_{\rm m0}/a^3$. This
provides $\rho_{\rm HH} \propto 1/a^3$, which implies $w_{\rm
m}=0=w_{\rm HH}$~\cite{HSU}. Furthermore, the first Friedmann
equation implies $\Omega_{\rm HH}+\Omega_{\rm m}=1$ with
$\Omega_{\rm HH}=c^2$. However, this is an unwanted case because
of $\Omega_{\rm m}={\rm const}$. Using the second Friedmann
equation (\ref{omee}), one has either $w_{\rm HH}=0$ or
$\Omega_{\rm HH}=1$. On the other hand, one may find from
Eq.(\ref{EOS})
\begin{equation} \label{rch}
w_{\rm HH}=-1-\frac{2\dot{H}}{3H^2} \end{equation} which  can be
rewritten as
\begin{equation}w_{\rm
HH}=-1+\frac{2a \epsilon}{3}\end{equation} with
$\epsilon=-\frac{\dot{H}}{aH^2}$.
 For $H \simeq {\rm
const}$, one finds $w_{\rm HH}=-1$. However, for $\epsilon >0$,
$w_{\rm HH}>-1$, while for $\epsilon <0$, $w_{\rm HH}<-1$. This
means that the holographic dark energy model with $L_{\rm HH}$
does not provide a promising EOS except the interacting
case~\cite{Horvat}.

For the particle horizon with $L_{\rm PH}=a\int^a_0 da'/a'^2H'$,
it could not describe the accelerating phase because of
\begin{equation}
w_{\rm PH}=-1+\frac{2}{3L_{\rm PH}}\frac{dL_{\rm
PH}}{dx}=-\frac{1}{3}+\frac{2 \sqrt{\Omega_{\rm PH}}}{3c} \ge
-1/3,~{\rm for}~ c \ge 1.
\end{equation}

 The only choice which
provides an accelerating phase is the future event horizon $L_{\rm
FH}=a\int^\infty_a da'/a'^2H'$ and thus its equation of state is
given by
\begin{equation}
w_{\rm FH}=-1+\frac{2}{3L_{\rm FH}}\frac{dL_{\rm
FH}}{dx}=-\frac{1}{3}-\frac{2 \sqrt{\Omega_{\rm FH}}}{3c} \le
-1/3,~{\rm for}~ c \ge 1. \end{equation}
 Hence, obtaining the accelerating phase is just the problem of
choice of the IR cutoff $L$  in the holographic dark energy
models. This is  because the logarithmic derivative of IR cutoff
is given by
\begin{equation} \label{crir}
\frac{dL_{\rm PH/FH}}{dx}=L_{\rm PH/FH}\pm \frac{1}{H}.
\end{equation}
 That is, from Eqs.(\ref{EOS}) and (\ref{crir}),
the rate of change for the size $L$ of universe determines the
equation of state within the holographic dark energy model. If $L$
is fixed, its EOS is $-1$ just like the cosmological constant.
Hence $L$ is rather ad hoc chosen.  In other words, there is no
such IR cutoff of future event horizon without first having a
holographic dark energy and there is no holographic dark energy
without first having an IR cutoff to define it.   This leads to a
conceptual paradox that is similar to the question of ``the
chicken and the egg"~\cite{Med}.

In order to understand this issue clearly, we introduce the
agegraphic and new agegraphic dark energy
densities~\cite{CAI,WC1,KLMage,KLMP}
\begin{equation}
\rho_{\rm T}=\frac{3c^2 m^2_{\rm p}}{T^2} ~~{\rm and}~~\rho_{\rm
\eta}=\frac{3c^2 m^2_{\rm p}}{\eta^2}
\end{equation}
with the same parameter $c$, respectively. At first sight, it
seems to  require the time scale of the universe.  We remind the
reader that we are working with the units of $c=\hbar=k_B=1$. In
this unit system, there is no essential difference between time
and length. Thus one may use the terms like time and length
interchangeably ($l=t$), where $l_{\rm P}=t_{\rm P}=1/m_{\rm P}$
being the reduced Planck length, time and mass, respectively. This
means that $\rho_{\rm T}$ and $\rho_{\rm \eta}$ are the same as
$\rho_{\rm \Lambda}$ except different IR cutoffs\footnote{For
example, we have the present age of the universe
$t_0=\int^{t_0}_0dt'$, the Hubble horizon
$H^{-1}_0=\frac{3}{2}t_0$, and the particle horizon $L^0_{\rm
PH}=a_0\int^{t_0}_0\frac{dt'}{a'}=3t_0$\cite{LL}. The distance
that the light travels is greater than we could get by naively
multiplying the age of universe by the speed of light.}. In this
sense, we choose the IR cutoff as
\begin{equation}
T=\int^t_0dt'=\int^x_{-\infty}\frac{dx'}{H'}
\end{equation}
which is the age of universe. In terms of length scale, it is the
logarithmic integral of the Hubble radius $H^{-1}$. In addition,
the conformal time is defined by
 \begin{equation}
\eta=\int^t_0\frac{dt'}{a'}=\int^x_{-\infty}\frac{dx'}{a'H'}\end{equation}
which  is the maximum comoving distance to a comoving observer's
particle horizon since $t=0$. That is, this is the logarithmic
integral of the comoving Hubble radius $1/aH$. We call it the
comoving horizon.

For the case that $H$ is nearly constant, one finds that these are
\begin{equation}
T=\int^a_{0}\frac{da'}{a'H'}\simeq \frac{\ln[a]}{H}\equiv T_{\rm
H},~\eta=\int^a_{0}\frac{da'}{a'^2H'}\simeq -\frac{1}{aH} \equiv
\eta_{\rm H}.\end{equation} In this case, we have approximate
forms of energy density
\begin{equation}
\tilde{\rho}_{\rm T} \simeq \frac{3c^2m^2_{\rm
p}H^2}{(\ln[a])^2},~\tilde{\rho}_{\rm \eta} \simeq 3c^2m^2_{\rm
p}a^2H^2,~\tilde{\rho}_{\rm PH/FH} \simeq 3c^2m^2_{\rm
p}H^2=\rho_{\rm HH},
\end{equation}
which shows that the energy densities of proper distance $L_{\rm
PH/FH}$ are approximately  the same as   that of the Hubble
horizon, $\rho_{\rm HH}$. This implies  that the holographic
energy density model with the proper distance  may be  regarded as
a ``dynamical cosmological constant model".

 The derivatives of these lead
to the same expression, respectively
\begin{equation}
\frac{dT}{dx}=\frac{dT_{\rm
H}}{dx}=\frac{1}{H},~~\frac{d\eta}{dx}=\frac{d\eta_{\rm
H}}{dx}=\frac{1}{aH}
\end{equation}
which shows that in calculating their EOS, there is no significant
difference even for choosing $H\simeq {\rm const}$. Namely, the
instantaneous rate of change for $T$ and $\eta$ are given by the
Hubble radius and comoving Hubble radius, respectively. Using
Eq.(\ref{EOS}), we have
\begin{equation}
w_{\rm  T}=-1+\frac{2\sqrt{\Omega_{\rm T}}}{3c},~w_{\rm \eta
}=-1+\frac{2e^{-x}\sqrt{\Omega_{\rm \eta}}}{3c}.
\end{equation}

On the other hand, we may introduce
\begin{equation}
\bar{T}=\int^\infty_tdt'=\int_a^{\infty}\frac{da'}{a'H'}=
\int_x^{\infty}\frac{dx'}{H'}\end{equation}
 to be the future age of
universe. Also
 \begin{equation}
 \bar{\eta}=\int^\infty_t\frac{dt'}{a'}=\int_a^{\infty}\frac{da'}{a'^2H'}=\int_x^{\infty}\frac{dx'}{a'H'}. \end{equation}
is the comoving distance to a comoving observer's future event
horizon. Then their derivatives are given by  opposite signs to
$T$ and $\eta$,
\begin{equation}
\frac{d\bar{T}}{dx}=-\frac{1}{H},~~\frac{d\bar{\eta}}{dx}=-\frac{1}{aH}.
\end{equation}
Their equations of state are given by
\begin{equation}
w_{\rm \bar{T}}=-1-\frac{2\sqrt{\Omega_{\rm T}}}{3c},~w_{\rm
\bar{\eta} }=-1-\frac{2e^{-x}\sqrt{\Omega_{\rm \eta}}}{3c}.
\end{equation}
We note that  $L_{\rm PH}=a \eta $ is the proper distance to
particle horizon, whereas $L_{\rm FH}=a \bar{\eta}$ is the proper
distance to future event horizon. $L_{\rm FH}$ is the distance to
the most distance event we will ever see (the distance light can
travel between now and the end of time) in contrast to $L_{\rm
PH}$, which is the distance to the most distant object we can
currently see (the distance light has travelled since the
beginning of time). An externally expanding model possesses future
event horizon if light can not travel more than a finite distance
in an infinite time, $\bar{\eta}<\infty$~\cite{CTS}. However, we
do not have the future event horizon for the future age of
universe because of $\bar{T}\sim \infty$. Thus we exclude this
case from our consideration.

 For the choice of proper
distance, we have non-accelerating phase for particle horizon,
while we have the accelerating phase for future event horizon with
$c\ge 1$. On the contrary to this, for the choice of coordinate
distance ($\eta$=comoving distance), we have accelerating phase
for particle horizon, while we have super-accelerating (phantom)
phase for future event horizon with any $c$.  This shows the
apparent difference between holographic and new agegraphic dark
energy models. However, there is no essential difference between
two models. The apparent difference is  to choose a different
distance.

The causality issue may be resolved for agegraphic and new
agegraphic dark energy model when choosing the coordinate
distance.  This is possible because the conformal time $\eta$ as
an IR cutoff exists in the new agegraphic dark energy model,
irrespective of the existence of the eternal accelerated expansion
in the future~\cite{WC1}. On the other hand,  this issue arises
for the holographic dark energy model with the proper distance.
This is because in order to have an accelerating universe, one
chooses the future event horizon which shows the eternal
accelerated expansion of the universe in the future. However, an
accelerating phase may arise as a pure interaction phenomenon if
pressureless dark matter is coupled to holographic dark energy
whose IR cutoff scale is set by the Hubble length~\cite{ZP}.

Finally, we would like to mention that the causality issue may be
not resolved for the new agegraphic dark energy model in the
future. The coordinate (comoving) distance induces more
acceleration than the proper distance. Actually, we observe that
$w_{\rm \eta} \to -1$, irrespective of $c$ in the
future~\cite{KLMage}. This implies the presence of the future
event horizon because the accelerating phase of $-1/3<w \le -1$
could  develop the future event horizon in the future~\cite{CTS}.

\section{Discussions}
The spacetime foam model  could provide  the holographic energy
density. However, its holographic model which implies the exotic
matter, a dark energy with infinite statistics is not a unique way
to derive the holographic energy density.

Furthermore, even if one gets the form of holographic energy
density, it is a separate issue to find an accelerating universe
from this density.  Hence we may choose IR cutoff to be  a
dynamical length scale like either  coordinate distance (age of
universe $T$ and comoving distance $\eta$) or proper distance
(particle horizon $L_{\rm PH}$ and future event horizon $L_{\rm
FH}$). The cases of comoving distance $\eta$ and proper distance
$L_{\rm FH}$ could explain an accelerating phase of the universe.
However, it is unclear which distance is appropriate for the
description of a dark-energy dominated universe. Along this
direction, the proper distance of particle horizon $L_{\rm PH}$
was used to calculate the entropy bound~\cite{FS,DPS}.

Until now, we do not know the nature of an exotic matter which may
derive an accelerating universe because  both of ordinary and
exotic matters could lead to the holographic energy density as
well as  it is a matter of choice of IR cutoff to obtain an
accelerating universe.

\section*{Acknowledgment}

This was  in part supported by  the Korea Research Foundation
(KRF-2006-311-C00249) funded by the Korea Government (MOEHRD), and
by the Science Research Center Program of the Korea Science and
Engineering Foundation through the Center for Quantum Spacetime of
Sogang University with grant number R11-2005-021.

        \end{document}